\documentclass[%
 aip,
 amsmath,amssymb,
reprint,%
]{revtex4-1}

\usepackage{graphicx}
\usepackage{dcolumn}
\usepackage{bm}

\usepackage[utf8]{inputenc}
\usepackage[T1]{fontenc}
\usepackage{mathptmx,textcomp,mathtools}

\begin{document}

\preprint{AIP/123-QED}

\title{Tunable ferromagnetic resonance in coupled trilayers with crossed in-plane and perpendicular magnetic anisotropies}

\author{Daniel Mark\'o}
 \email{daniel.marko@uvsq.fr}
	\affiliation{Groupe d’Etude de la Mati\`ere Condens\'ee (GEMaC), UMR8635, CNRS and Universit\'e de Versailles/Saint-Quentin-en-Yvelines, Universit\'e Paris-Saclay, 45 ave des Etats-Unis, 78035 Versailles, France}
\author{Fernando Vald\'es-Bango}%
 \affiliation{Departamento de F\'isica, Facultad de Ciencias, Universidad de Oviedo, C/ Federico Garc\'ia Lorca n$^\circ$ 18, 33007 Oviedo, Spain}
\author{Carlos Quir\'os}%
 \affiliation{Departamento de F\'isica, Facultad de Ciencias, Universidad de Oviedo, C/ Federico Garc\'ia Lorca n$^\circ$ 18, 33007 Oviedo, Spain}
	\affiliation{Centro de Investigaci\'on en Nanomateriales y Nanotecnolog\'ia (CINN), CSIC-Universidad de Oviedo, Spain}
\author{Aurelio Hierro-Rodr\'iguez}%
 \affiliation{SUPA, University of Glasgow, School of Physics and Astronomy, G128QQ Glasgow, UK}
\author{Mar\'ia V\'elez}%
 \affiliation{Departamento de F\'isica, Facultad de Ciencias, Universidad de Oviedo, C/ Federico Garc\'ia Lorca n$^\circ$ 18, 33007 Oviedo, Spain}
	\affiliation{Centro de Investigaci\'on en Nanomateriales y Nanotecnolog\'ia (CINN), CSIC-Universidad de Oviedo, Spain}
\author{Jos\'e Ignacio Mart\'in}%
 \affiliation{Departamento de F\'isica, Facultad de Ciencias, Universidad de Oviedo, C/ Federico Garc\'ia Lorca n$^\circ$ 18, 33007 Oviedo, Spain}
	\affiliation{Centro de Investigaci\'on en Nanomateriales y Nanotecnolog\'ia (CINN), CSIC-Universidad de Oviedo, Spain}
\author{Jos\'e Mar\'ia Alameda}%
 \affiliation{Departamento de F\'isica, Facultad de Ciencias, Universidad de Oviedo, C/ Federico Garc\'ia Lorca n$^\circ$ 18, 33007 Oviedo, Spain}
	\affiliation{Centro de Investigaci\'on en Nanomateriales y Nanotecnolog\'ia (CINN), CSIC-Universidad de Oviedo, Spain}
\author{David S. Schmool}%
 \affiliation{Groupe d’Etude de la Mati\`ere Condens\'ee (GEMaC), UMR8635, CNRS and Universit\'e de Versailles/Saint-Quentin-en-Yvelines, Universit\'e Paris-Saclay, 45 ave des Etats-Unis, 78035 Versailles, France}
\author{Luis Manuel \'Alvarez-Prado}%
 \email{lmap@uniovi.es}
	\affiliation{Departamento de F\'isica, Facultad de Ciencias, Universidad de Oviedo, C/ Federico Garc\'ia Lorca n$^\circ$ 18, 33007 Oviedo, Spain}
	 \affiliation{Centro de Investigaci\'on en Nanomateriales y Nanotecnolog\'ia (CINN), CSIC-Universidad de Oviedo, Spain}

\date{\today}

\begin{abstract}
A novel approach to tune the ferromagnetic resonance frequency of a soft magnetic Ni$_\text{80}$Fe$_\text{20}$ (Permalloy = Py) film with in-plane magnetic anisotropy (IMA) based on the controlled coupling to a 
hard magnetic NdCo$_\text{x}$ film with perpendicular magnetic anisotropy (PMA) through a non-magnetic Al spacer is studied. Using transverse magneto-optical Kerr effect (TMOKE), alternating gradient magnetometry
(AGM) as well as vector network analyzer ferromagnetic resonance (VNA-FMR) spectroscopy, the influence of both Co concentration and Al spacer thickness on the static and dynamic magnetic properties 
of the coupled IMA/PMA system is investigated. Compared to a single Py film, two striking effects of the coupling between IMA and PMA layers can be observed in their FMR spectra. First, there is a significant 
increase in the zero-field resonance frequency from 1.3 GHz up to 6.6 GHz, and second, an additional frequency hysteresis occurs at low magnetic fields applied along the hard axis. The maximum frequency difference between the frequency branches for increasing and decreasing magnetic field is as high as 1 GHz, corresponding to a tunability of about 20\% at external fields of typically less than $\pm$70 mT. The origin of the observed features in the FMR spectra is discussed by means of magnetization reversal curves. 
\end{abstract}

\maketitle
The magnetic properties of thin films and multilayers exhibiting stripe domains have been investigated extensively in both experiment and theory since their discovery more than half a century ago 
\cite{Fujiwara1964}. In recent years, research results on stripe domains have triggered the prospect of employing their unique properties in future microwave, magnonic, and spintronic devices with novel 
functionalities. The formation of stripe domains is the result of energy minimization as well as the competition between PMA ($K_\perp$) and shape anisotropy ($\frac{1}{2}\mu_0M_S^2$), which favor out-of-plane and 
in-plane magnetization, respectively. The ratio $Q = 2K_\perp/\mu_0M_S^2$, known as reduced anisotropy or quality factor \cite{HS1998}, is commonly used to describe the extent of stripe domains. For moderate 
($Q\!<\!1$) to weak ($Q\!\ll\!1$) PMA, the magnetization tends to lie in the plane, but above a critical film thickness $d_\text{cr}$, a ground state with stripe domains emerges. The latter is characterized by a 
perpendicular  magnetization component alternating between up and down within a period $\lambda$. The critical thickness $d_\text{cr}$ is typically in the range of 20 – 40 nm for moderate $Q$ value materials such as 
amorphous NdCo alloys \cite{Hierro-Rodriguez2012,Hierro-Rodriguez2013}, whereas for materials like Py with small values of $Q$, generally larger values of $d_\text{cr}$ = 170 – 300 nm are found 
\cite{BenYoussef2004,Ramos2009,TeeSoh2013,Wei2015}. Intimately linked to the presence of stripe domains is the occurrence of a pseudo-uniaxial or rotatable anisotropy \cite{LMAP97,Tacchi2014}, which is the result of 
the in-plane magnetization being aligned along the stripe direction. The latter, however, is not fixed as it can be reoriented by applying a saturating field along an arbitrary in-plane direction. This particular 
property of stripe domains has been shown recently to enable tunable and reconfigurable dynamic magnetic properties \cite{Banerjee2017,Pan2018} even after sample preparation and hence in contrast to other 
approaches of increasing FMR and spin wave frequencies in soft magnetic thin films \cite{Chai2012,Li2015,Li2016}. Another possibility to create stripe domains in soft magnetic materials even far below the critical 
thickness $d_\text{cr}$ stems from the coupling to another magnetic thin film or multilayer stack exhibiting PMA. Here, the influence of stray field and exchange interaction on the soft magnetic layer has been 
shown to lead to a multitude of intriguing effects such as, for example, imprinted topological spin textures \cite{Gilbert2015,Blanco-Roldan2015,Hierro-Rodriguez2017PRB}, deterministic 
propagation of vortex-antivortex pairs \cite{Hierro-Rodriguez2017}, and spin wave propagation in domain wall-like magnetic channels \cite{Wang2018}. Though magnetization dynamics of stripe domains in uncoupled thin 
films has been studied extensively \cite{Acher1997,Ebels2001,Vukadinovic2001,Buznikov2002,Acher2003,BenYoussef2004,Ramos2009,Ha2009,Chai2013,TeeSoh2013,Tacchi2014,Wei2015,Wei2016}, the dynamic magnetic properties 
of coupled IMA/PMA systems have so far only been investigated in a few studies \cite{Teixeira2016,Fallarino2016,Xu2018}.
\begin{figure*}
	\centering
		\includegraphics[width=0.99\textwidth]{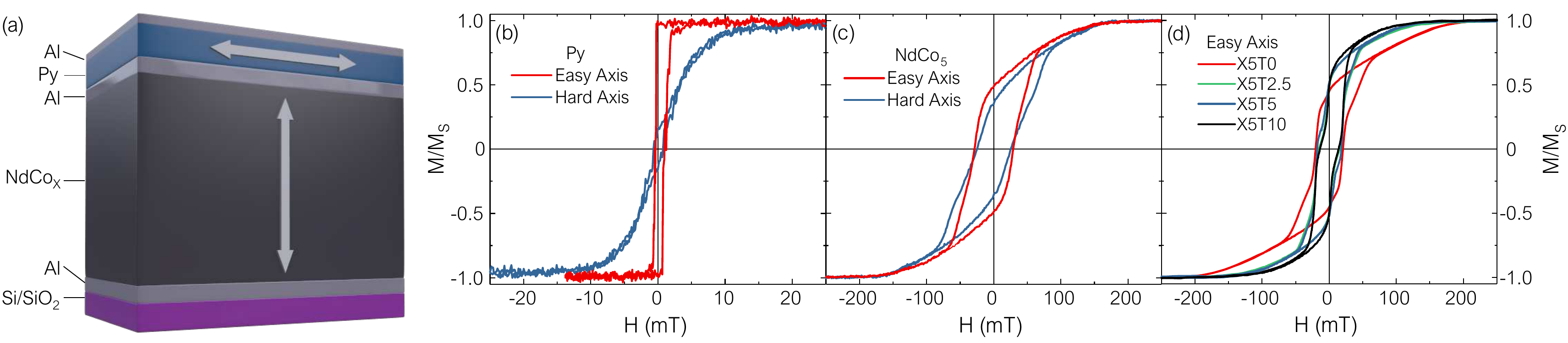}
\caption{(Color online) (a) Sketch of a coupled trilayer with arrows indicating the anisotropy directions in the magnetic films. (b) In-plane EA and HA hysteresis loops of a single 10 nm thick Py film 
					measured by T-MOKE. (c) In-plane EA and HA hysteresis loops of a single 64 nm thick NdCo$_5$ film measured by AGM. (d) In-plane EA hysteresis loops of X5 series samples for all 
					values of the Al spacer thickness $T$ measured by AGM.}
	\label{fig:Fig1}
\end{figure*}

In this Letter, a novel approach to tune the FMR frequency of a soft magnetic thin film based on the controlled coupling of two magnetic films with different types of anisotropies, in-plane and perpendicular, is 
investigated experimentally. Making use of the stripe domains' unique dynamic properties, a reconfigurable FMR response at low magnetic fields has been achieved. A 
sketch of the samples fabricated for this work is shown in Fig.~\ref{fig:Fig1}(a). The central element is a trilayer consisting of a 64 nm thick amorphous NdCo$_\text{x}$ film with PMA and a 10 nm thick 
polycrystalline Py film with IMA, which are coupled through a non-magnetic Al spacer. The trilayer structure itself is sandwiched between Al seed and capping layers, all of which have been grown on a Si/SiO$_2$ 
substrate using magnetron sputtering. The magnetic properties of the coupled thin films can be controlled by two independent parameters. On the one hand, varying the Co concentration ($X = {5, 7.5, 9}$) in the NdCo$_
\text{x}$ film allows the modification of the strength of its PMA. A maximum has been found for $X=5$, whereas higher or lower Co concentrations lead to a gradually weaker PMA, respectively 
\cite{Mergel1993,Cid2017}. On the other hand, by adjusting the Al spacer thickness ($T$ = {0 nm, 2.5 nm, 5 nm, 10 nm}), the type of coupling between the two magnetic layers can be set to either direct exchange 
coupling ($T\!\leq\!1.5$ nm) or stray field coupling ($T\!\geq\!2.5$ nm). In addition to the coupled bi- and trilayers, a series of reference samples, consisting of a single 10 nm thick Py film as well as single 64 
nm thick NdCo$_\text{x}$ films with varying Co concentrations, have also been prepared. For the remainder of the paper, the coupled bi- and trilayers will be named according to their Co concentration and Al spacer 
thickness as, e.g., X5T10 for a sample based on a NdCo$_5$ film and a 10 nm thick Al spacer.  

The static magnetic properties of the samples have been investigated using T-MOKE and AGM. In Fig.~\ref{fig:Fig1}(b) and (c), both in-plane easy axis (EA) and hard axis (HA) hysteresis loops of single Py and NdCo$_\text{5}$ films obtained by T-MOKE and AGM, respectively, are shown. The magnetization reversal loops of the Py film show the typical features of a soft magnetic material such as very low coercivity, low saturation field, and, for the EA, almost perfect squareness of the loop. In contrast, the hysteresis loops of the NdCo$_5$ film show a much larger coercivity and a higher saturation field, as expected for a high-anisotropy material. The reason for the higher in-plane remanence $M_\text{r}$ of the NdCo$_5$ loop in the EA configuration is a smaller out-of-plane component of the magnetization compared to the HA configuration. Upon coupling these two magnetic films to form either bilayers (without Al spacer) or trilayers (with Al spacer of variable thickness) with crossed anisotropies, respectively, the resulting magnetic properties are different from those of the individual films, yet they do not simply constitute a superposition or averaging due to the magnetic coupling between the layers. As an example, in-plane EA hysteresis loops of the X5 sample series for all four values of the Al spacer thickness measured by TMOKE are depicted in Fig.~\ref{fig:Fig1}(d). For the bilayer system with direct exchange coupling due to a very thin Al spacer, the resulting hysteresis loop is very similar to that of the single NdCo$_5$ film. This indicates that the Py layer effectively behaves like the NdCo$_5$ film and can be considered almost as an extension of the hard magnetic layer. However, for increasing Al spacer thicknesses, the magnetic coupling reduces, meaning the Py acts more and more as a soft magnetic film, which is important for its dynamic behavior. 
\begin{figure*}
  \begin{minipage}[c]{0.69\textwidth}
    \includegraphics[width=\textwidth]{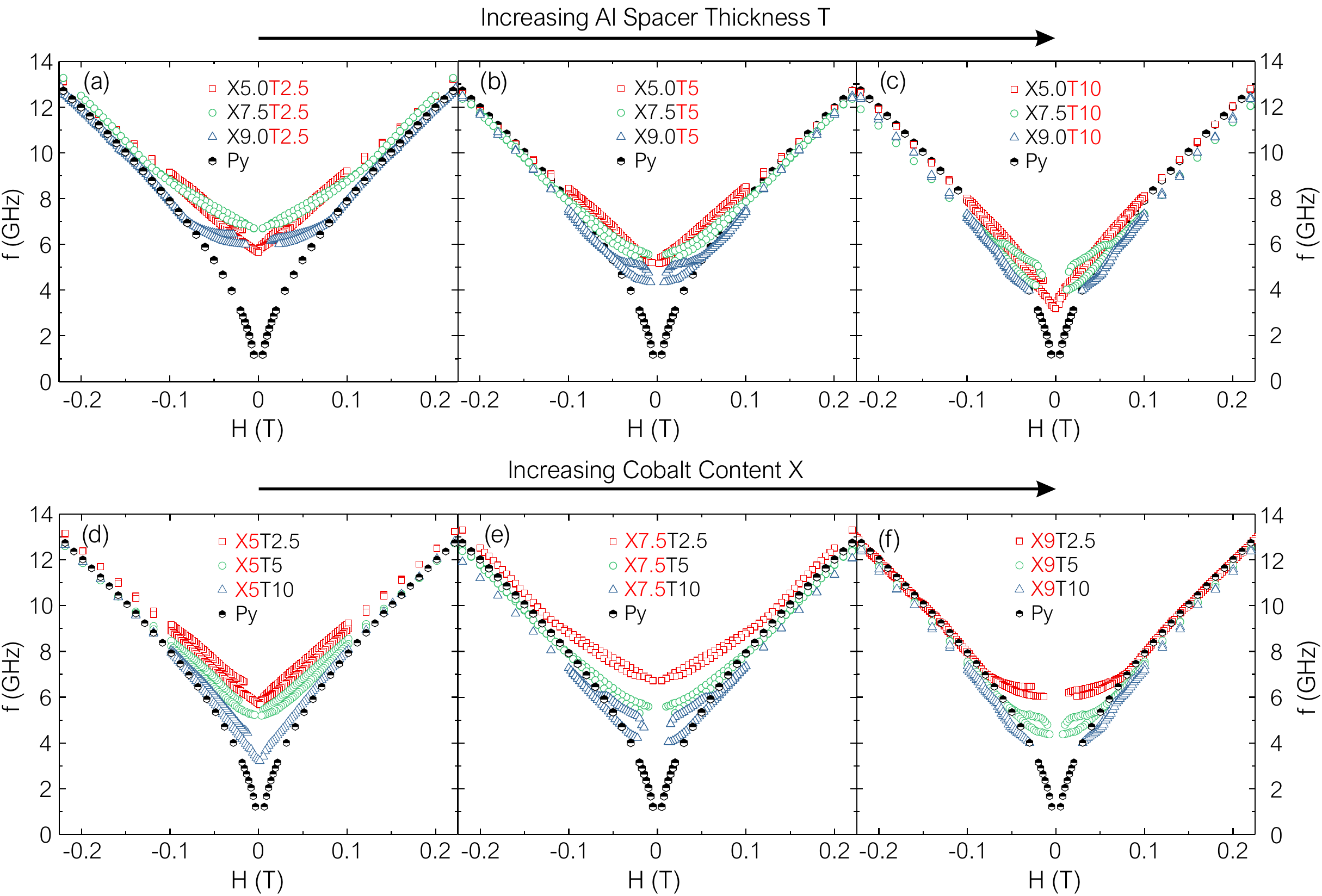}
  \end{minipage}\hfill
  \begin{minipage}[c]{0.29\textwidth}
    \caption{(Color online) $f$ vs.~$H$ dependency of the coupled trilayers for the in-plane dc magnetic field $H$ applied long the HA. In the top row (a$-$c), the FMR spectra of trilayers having the same Al 
					spacer thickness ($T$) are compared, whereas in the bottom row (d$-$f), the results for trilayers with identical Co concentrations ($X$) are displayed. In both rows, the values of $X$ and $T$ increase 
					from left to right, respectively. For comparison, the spectrum of a single 10 nm thick Py film has been added to each plot. Prior to each measurement, stripe domains and magnetization have been aligned 	
					parallel by the application and removal of a positive in-plane saturating magnetic field, which was then swept from 0 T $ \rightarrow$ 0.9 T $\rightarrow$ $-$0.9 T $\rightarrow$ 0 T during the actual FMR 
					experiment.}
		\label{fig:Fig2}
	\end{minipage}
\end{figure*}
The dynamic magnetic properties of the samples have been investigated by means of room temperature broadband VNA-FMR using the flip-chip method, in which the sample is placed upside-down on top of a coplanar waveguide with a 50 \textmu m wide center conductor. The VNA was operated in frequency sweep mode while an in-plane dc magnetic field $H$, applied either along the EA or  HA of the samples, was swept in the following sequence: 0 T $\rightarrow$ 0.9 T $\rightarrow$ $-$0.9 T $\rightarrow$ 0 T. Prior to each measurement, the samples were saturated in order to ensure that at the beginning of the actual FMR experiment the dc magnetic field $H$ is parallel and the rf magnetic field $H_\text{rf}$, generated by the CPW, is perpendicular to the stripe domains, respectively. The magnitude of the forward transmission parameter $S_{21}$ was used to extract the resonance frequencies $f$ after a reference spectrum taken at zero-field was subtracted from all of the recorded spectra. In Fig.~\ref{fig:Fig2}, the
$f$ vs.~$H$ dependency for the dc magnetic field applied along the HA of the coupled trilayers is displayed. In the upper row (a$-$c), the thickness of the Al spacer $T$ increases from the left to the right panel, while the Co concentration $X$ varies for every fixed value of $T$ in each of the panels. Accordingly, in the bottom row (d$-$f), the Co concentration of the 
NdCo$_\text{x}$ films increases from the left to the right panel, while the Al spacer thickness $T$ varies for every fixed value of $X$. As such, the same data is shown in (a$-$c) and (d$-$f). For comparison, the FMR spectrum of a single 10 nm thick Py reference film has been included in all panels. Due to the phenomenological damping and the corresponding large linewidth, it was not possible to extract any data from the FMR spectra of any single NdCo$_\text{x}$ film as well as any of the directly exchange-coupled bilayers. However, the insertion of the Al spacer with its variable thickness $T$ leads to a gradual decoupling of the IMA/PMA stack, thereby effectively enabling the observation of the FMR of the soft Py film, whose magnetic properties are modified by the proximity to the hard NdCo$_\text{x}$ layer, resulting in both an induced rotatable anisotropy and a stripe domain pattern. All FMR spectra in Fig.~\ref{fig:Fig2} show exactly one single resonance: either the uniform FMR mode in the case of the single Py film or an acoustic mode in the coupled trilayers, which becomes the uniform mode when the samples are saturated and the stripe domains are erased. The origin of the acoustic mode is the in-phase precession of spins in adjacent stripe domains. At lower fields, where the stripe domains in the coupled IMA/PMA samples are nucleated, a significant deviation from the single Py frequencies can be seen, which manifests itself by two very distinct features. First, there is a strong increase in the zero-field resonance frequencies from about 1.3 GHz for Py up to a maximum of 6.6 GHz for the X7.5T2.5 trilayer and second, there is also a frequency hysteresis with differences between the two field sweep directions as high as 1 GHz in the case of the X7.5T10 trilayer. Within the hysteretic part of the FMR spectra, the lower frequency branch at negative fields and the higher frequency branch at positive fields can be accessed when increasing the value of the applied magnetic field. Conversely, the lower frequency branch at positive fields and the higher frequency branch at negative fields can be accessed when decreasing the value of the applied magnetic field. Although the hysteretic behavior of the $f$ vs.~$H$ dependency is a rather rare phenomenon, it has been observed in a variety of materials including, e.g., exchange-biased bilayers \cite{Spenato2005}, BaFe$_{12}$O$_{9}$ films \cite{Wang2002,Kostenko2008}, thick Py films \cite{Acher2003,Cao2019}, artificial spin ice \cite{Jungfleisch2016}, and patterned nanostructures based on Py \cite{Giesen2005,Nozaki2008}. This effect allows the resonance frequency to be tuned as a function of the magnetic history, leading to a reconfigurable functionality in a Py film exhibiting stripe domains at a thickness of just 10 nm. From the top panels (a$-$c) in Fig.~\ref{fig:Fig2}, in which the results for samples with fixed Al spacer thickness are shown, it can be seen that an increase of the Co concentration $X$ in the NdCo$_\text{x}$ alloys leads to a decrease of the resonance frequencies due to its reduced PMA, resulting in a gradual convergence of the frequencies within the hysteretic part of the spectra to the frequencies of the single Py film. Similarly, as depicted in the lower panels (d$-$f) in Fig.~\ref{fig:Fig2}, an increase of the Al spacer thickness for a constant Co concentration leads to a decrease of the FMR frequencies and their gradual convergence towards the single Py film frequencies. The reason for this is the weaker influence of the NdCo$_\text{x}$ stray field on the Py film with increasing distance between both these two films.

\begin{figure}
 \centering
  \includegraphics[width=0.44\textwidth]{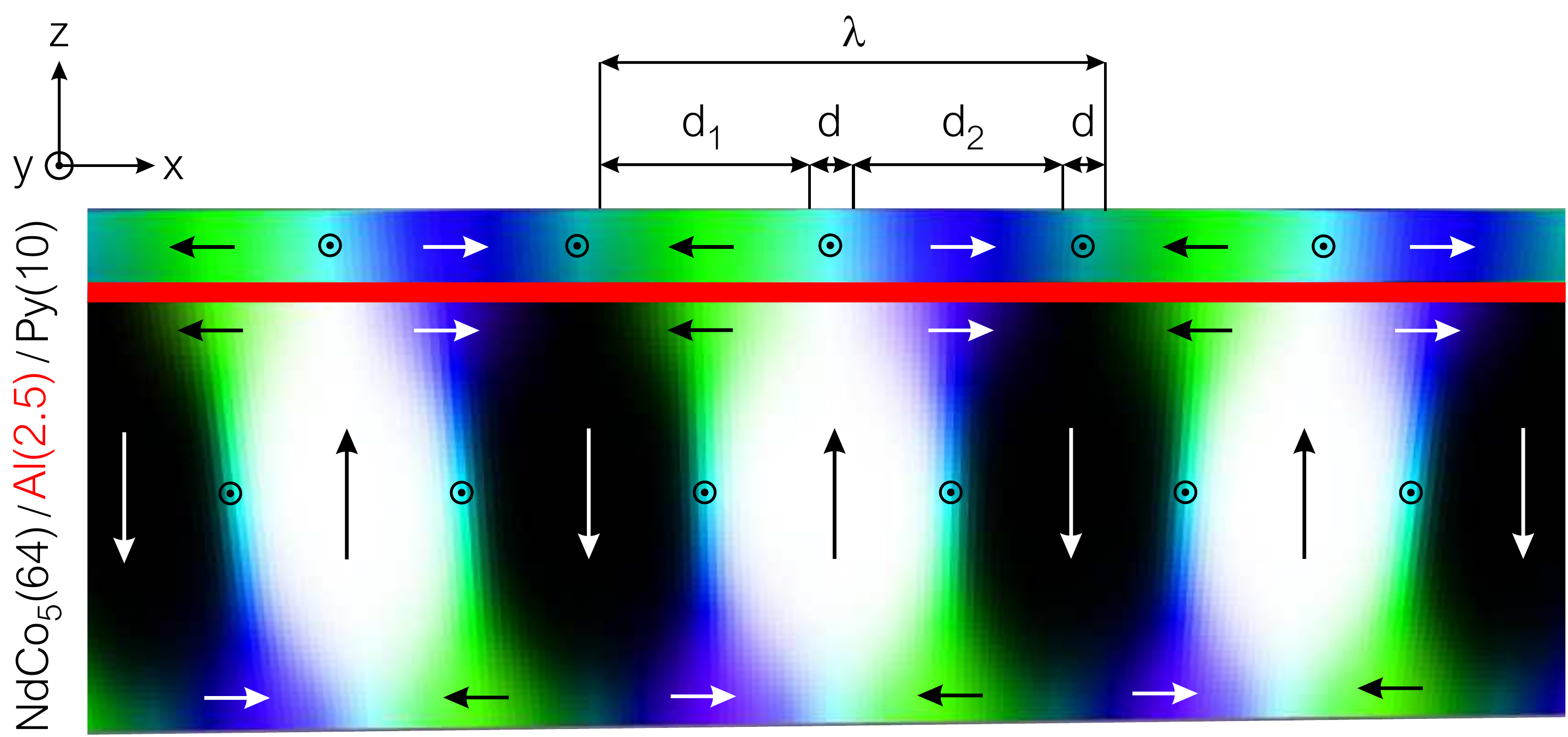}
	 \caption{(Color online) Simulated magnetic domain configuration in a X5T2.5 trilayer at remanence after application of a saturating magnetic field along the $y$-direction. The stray field of the stripe domains 
						in the NdCo$_5$ layer creates a closure domain pattern in the Py layer even across the thin non-magnetic Al spacer.}
	\label{fig:Fig4}
\end{figure}

In Fig.~\ref{fig:Fig4}, the simulated stripe domain pattern in a X5T2.5 trilayer at remanence after saturation with a magnetic field applied along the $y$-direction is depicted. In the NdCo$_5$ layer, 
$m_\text{z}$ is alternatingly pointing up or down, forming stripe domains of periodicity $\lambda$ that are separated by Bloch walls in which $m_\text{y}$ is maximum. In order to minimize the stray field energy, the $x$-component of the magnetization forms closure domains, indicated by black/white arrows pointing left/right, at both top and bottom of the NdCo$_5$ layer. This closure domain pattern is also imprinted and hence extended across the thin Al spacer into the Py layer, where regions with opposite values of $m_\text{x}$ are separated by N\'eel walls in which $m_\text{y}$ is maximum. The replication of the weak stripe pattern in the Py layer also leads to a transfer of the rotatable anisotropy, allowing the Py film in the coupled trilayers to have a much larger IMA than a single Py film. Moreover, it is interesting to note that the lines of maximum $m_\text{y}$ within the Bloch walls in the PMA layer are shifted by $\lambda$/4 with respect to the ones within the N\'eel walls in the IMA layer. In addition, it can be seen that the stripe domain periodicity $\lambda$ in the Py layer is given by $d_1+d_2+2d$, i.e., the sum of the width of two closure domains (d$_{1,2}$) with opposite magnetization ($M_{1,2}$) as well as the width of two N\'eel walls (2$d$) separating them. Typical values of $\lambda$ for single NdCo$_\text{x}$ films as well as coupled IMA/PMA samples are in the range from 145$-$180 nm and 130$-$145 nm, respectively, as determined from magnetic force microscopy images.

\begin{figure}
	\centering
		\includegraphics[width=0.44\textwidth]{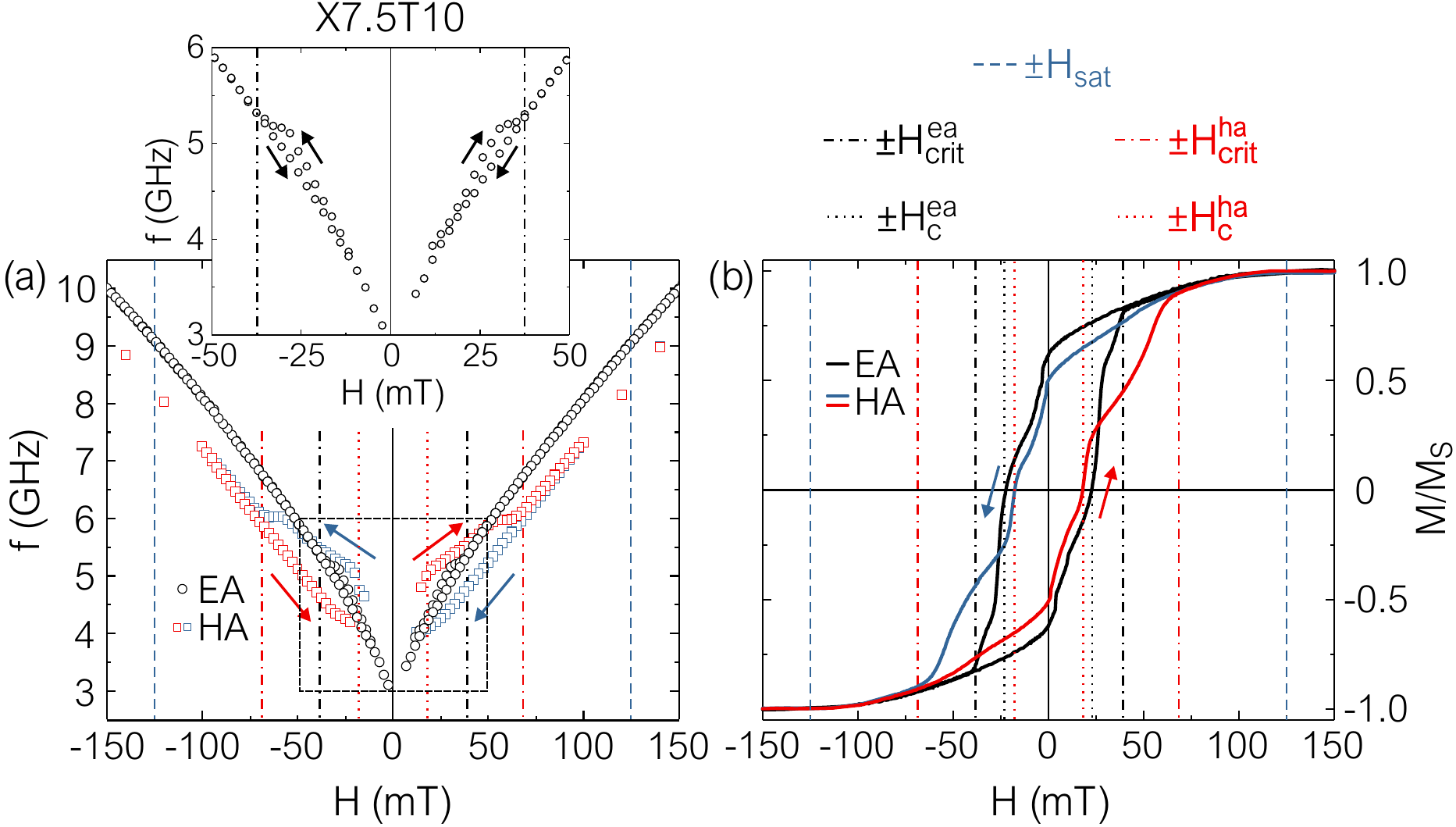}
	\caption{(Color online) In-plane EA and HA FMR spectra (a) and magnetization reversal curves measured by AGM (b) of the X7.5T10 trilayer. The arrows designate the field sweep directions, whereas the 
		dashed, dashdotted, and dotted lines indicate saturation field ($H_\text{sat}$), critical fields ($H_\text{crit}$), and coercive field ($H_\text{c}$), respectively. The small inset at the top of (a) is a zoom 
		into the area bound by the dashed rectangle to better see the minor frequency hysteresis in the EA configuration.}
	\label{fig:Fig5}
\end{figure}

In the following, a possible explanation for the observed frequency hysteresis will be discussed using the FMR spectra and magnetization reversal curves of the X7.5T10 trilayer measured with the magnetic 
field $H$ applied along EA and HA, respectively. In both sets of data, depicted in Fig.~\ref{fig:Fig5}(a) and (b), respectively, three characteristic fields can be identified, whose values 
are in excellent agreement. Those are the saturation field $H_\text{sat}$, the coercive field $H_\text{c}$, as well as the critical field $H_\text{crit}$, at which the splitting/merging of the hysteresis loop and 
FMR frequency branches occurs. While there is a sizeable frequency hysteresis when $H$ is applied along the HA, with the maximum frequency difference between both field sweep directions occurring at $H_\text{c}$, 
there is typically no or only a much less pronounced frequency hysteresis observed when $H$ is applied along the EA. The two hysteresis loops shown in Fig.~\ref{fig:Fig5}(b) differ in three important points: the EA 
loop has (i) a 20\% higher coercivity $H_\text{c}$, (ii) a 20\% higher remanence $M_\text{r}$, but (iii) an almost 50\% lower critical field $H_\text{crit}$ compared to the HA loop. 

For FMR measurements, the influence of different relative orientations of stripe domains and rf magnetic field $H_\text{rf}$ has been shown to lead to higher (lower) resonance frequencies in case of parallel 
(perpendicular) alignment as a result of the excitation of optical (acoustic) modes due to out-of-phase (in-phase) precession of the magnetization in adjacent stripe domains 
\cite{Ebels2001,Vukadinovic2001,Tacchi2014,Wei2015,Pan2018,Cao2019}. However, the way the FMR measurements in this work have been performed, stripe domains and rf magnetic field $H_\text{rf}$ (dc magnetic field $H$)
are always perpendicular (parallel) during the entire hysteresis cycle and independent of the field sweep direction as simulated in Ref.~[39] for a single 200 nm thick Py film. This means that both 
frequency branches in the HA FMR spectra of the coupled trilayers at low fields always correspond to an acoustic mode. Instead, the fact that generally no or only a minor frequency hysteresis can be observed in 
the EA configuration suggests that the IMA of both the Py and NdCo$_\text{x}$ layer and their relative orientation with respect to the in-plane dc magnetic field $H$ are at the origin of the observed dynamic 
properties. Co-sputtering generally induces an IMA in the NdCo$_\text{x}$ films of around 10$^4$ J/m$^3$, which is about one order of magnitude larger than the IMA of the Py layer even after rescaling the energy 
density with the corresponding values of $M_\text{S}$. Moreover, the IMA in the NdCo$_\text{x}$ alloys creates a huge asymmetry in the closure domain structure when the stripes are oriented along EA or HA. Thus, 
there is a relevant difference between the stray fields generated by the NdCo$_\text{x}$ film and the Py layer, respectively, depending on the relative directions of the IMA and the magnetization components of the 
closure/stripe domains. However, to gain further insight into this complex interplay, additional measurements of the azimuthal angle dependency of the FMR are necessary to quantify the value of the IMA and, in particular, its rotatable anisotropy contribution.    
  
In summary, a novel approach to boost the FMR frequency of a soft magnetic Py film with IMA based on the controlled coupling to a hard magnetic NdCo$_\text{x}$ film with PMA through a non-magnetic Al spacer of 
variable thickness has been investigated experimentally. The two most striking effects observed, compared to a single Py film, are a significant increase in the zero-field FMR frequency from 1.3 GHz up to 6.6 GHz, 
and a frequency hysteresis at HA fields below $\pm$70 mT with a difference between the frequency branches for increasing and decreasing field of up to 1 GHz, both of which can clearly be attributed to the imprinted 
stripe domain pattern in the Py layer below saturation. The possibility to control anisotropy and coupling strength in this IMA/PMA system by adjusting the Co concentration in the PMA film and the Al spacer 
thickness, respectively, during sample fabrication allows the system to be predefined with respect to the value of the zero-field resonance frequency. In addition, the FMR frequencies can further be tuned and 
reconfigured by  simply erasing and nucleating a stripe domain pattern in the Py layer upon application of an in-plane magnetic field along its HA, opening new perspectives for the development of future microwave, 
spintronic or magnonic devices.

D.~S.~S.~and L.~M.~Á.-P.~acknowledge CNRS for financial support. A.~H.-R. acknowledges European Union's Horizon 2020 research and innovation program under the Marie Sk\l odowska-Curie Action 
H2020-MSCA-IF-2016-746958. This work is supported by Spanish MINECO under project FIS2016-76058 (AEI/FEDER, UE). 

\bibliography{OviedoAPLwithoutTitles}

\end{document}


\preprint{AIP/123-QED}

\title{Supplement of "Tunable ferromagnetic resonance in coupled trilayers with crossed in-plane and perpendicular magnetic anisotropies"}

\author{Daniel Mark\'o}
 	\affiliation{Groupe d’Etude de la Mati\`ere Condens\'ee (GEMaC), UMR8635, CNRS and Universit\'e de Versailles/Saint-Quentin-en-Yvelines, Universit\'e Paris-Saclay, 45 ave des Etats-Unis, 78035 Versailles, France}
\author{Fernando Vald\'es-Bango}%
 \affiliation{Departamento de F\'isica, Facultad de Ciencias, Universidad de Oviedo, C/ Federico Garc\'ia Lorca n$^\circ$ 18, 33007 Oviedo, Spain}
\author{Carlos Quir\'os}%
 \affiliation{Departamento de F\'isica, Facultad de Ciencias, Universidad de Oviedo, C/ Federico Garc\'ia Lorca n$^\circ$ 18, 33007 Oviedo, Spain}
	\affiliation{Centro de Investigaci\'on en Nanomateriales y Nanotecnolog\'ia (CINN), CSIC-Universidad de Oviedo, Spain}
\author{Aurelio Hierro-Rodr\'iguez}%
 \affiliation{SUPA, University of Glasgow, School of Physics and Astronomy, G128QQ Glasgow, UK}
\author{Mar\'ia V\'elez}%
 \affiliation{Departamento de F\'isica, Facultad de Ciencias, Universidad de Oviedo, C/ Federico Garc\'ia Lorca n$^\circ$ 18, 33007 Oviedo, Spain}
	\affiliation{Centro de Investigaci\'on en Nanomateriales y Nanotecnolog\'ia (CINN), CSIC-Universidad de Oviedo, Spain}
\author{Jos\'e Ignacio Mart\'in}%
 \affiliation{Departamento de F\'isica, Facultad de Ciencias, Universidad de Oviedo, C/ Federico Garc\'ia Lorca n$^\circ$ 18, 33007 Oviedo, Spain}
	\affiliation{Centro de Investigaci\'on en Nanomateriales y Nanotecnolog\'ia (CINN), CSIC-Universidad de Oviedo, Spain}
\author{Jos\'e Mar\'ia Alameda}%
 \affiliation{Departamento de F\'isica, Facultad de Ciencias, Universidad de Oviedo, C/ Federico Garc\'ia Lorca n$^\circ$ 18, 33007 Oviedo, Spain}
	\affiliation{Centro de Investigaci\'on en Nanomateriales y Nanotecnolog\'ia (CINN), CSIC-Universidad de Oviedo, Spain}
\author{David S. Schmool}%
 \affiliation{Groupe d’Etude de la Mati\`ere Condens\'ee (GEMaC), UMR8635, CNRS and Universit\'e de Versailles/Saint-Quentin-en-Yvelines, Universit\'e Paris-Saclay, 45 ave des Etats-Unis, 78035 Versailles, France}
\author{Luis Manuel \'Alvarez-Prado}%
 	\affiliation{Departamento de F\'isica, Facultad de Ciencias, Universidad de Oviedo, C/ Federico Garc\'ia Lorca n$^\circ$ 18, 33007 Oviedo, Spain}
	 \affiliation{Centro de Investigaci\'on en Nanomateriales y Nanotecnolog\'ia (CINN), CSIC-Universidad de Oviedo, Spain}

\date{\today}

\maketitle
\section{Magnetic Force Microscopy}
\begin{figure}[hb]
 \centering
  \includegraphics[width=0.49\textwidth]{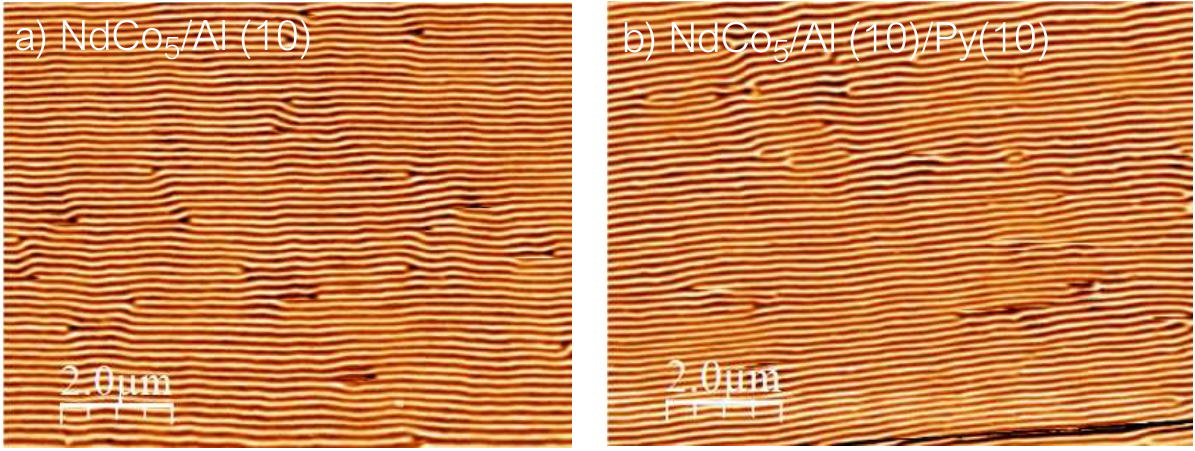}
	 \caption{(Color online) MFM images of a X5T10 sample (a) before and (b) after the deposition of the 10 nm thin Py layer. The stripe domain pattern of the NdCo$_5$ film is replicated in the Py layer of the PMA/IMA trilayer with only a minor difference in the periodicity: $\lambda$ = 145 nm in (a) and $\lambda$ = 140 nm in (b) for the dc magnetic field applied along the HA.}
	\label{fig:Fig6MFM}
\end{figure}

The magnetic stripe domain structure of the samples has been imaged by means of magnetic force microscopy (MFM) as described in Ref.~[\cite{Hierro-Rodriguez2012}] and the images taken were used to determine the 
corresponding stripe domain periodicity $\lambda$. For each samples series, X5 to X9, MFM images were taken both before and after the deposition of the Py layer and for the in-plane dc magnetic field applied along 
EA and HA, respectively. As an example, the stripe domains in a X5T10 sample before and after the deposition of the Py film are shown in Fig.~\ref{fig:Fig6MFM}(a) and (b), respectively. As result of the coupling 
between the soft and hard magnetic layers, the stripe domain pattern of the NdCo$_5$ film is replicated in the Py film with only a small difference in $\lambda$: 145 nm in the PMA film and 140 nm in the PMA/IMA 
trilayer for a dc magnetic field applied along the HA of the samples.

\section{Micromagnetic Simulations} 
Micromagnetic simulations using the MuMax3 code \cite{Vansteenkiste2014} have been performed with the purpose of calculating the FMR frequencies of the Py layer and simulating the magnetic domain configuration of 
the coupled trilayers. The following set of material parameters was used for the simulations: $\gamma = 15.9\times10^{10}$ rad/T\:s, NdCo$_5$: $K_\perp = 16.2\times10^4$ J/m$^3$, $M_\text{S}$ = 10$^6$ A/m, 
$A$ = 0.7$\times$10$^{-11}$ J/m, Py: $K_\text{u}$ = 423 J/m$^3$, $M_\text{S}$ = 846$\times$10$^3$ A/m, $A$ = 1.2$\times$10$^{-11}$ J/m, where $\gamma$ is the gyromagnetic ratio, $K_\perp$ the PMA, $M_\text{S}$ the 
saturation magnetization, $A$ the exchange stiffness constant, and $K_\text{u}$ the uniaxial IMA. The trilayer structures were modelled by cells with lateral dimensions of 2.5 nm in-plane as well as 1.25 nm, 1 nm 
and 2 nm out-of-plane for an aluminum spacer thickness $T$ of 2.5 nm, 5 nm, and 10 nm, respectively. The dynamic response of the trilayers was micromagnetically simulated as follows: Once the steady state 
of magnetization has been reached, a magnetic field pulse is applied perpendicular to the in-plane dc magnetic field $H$. The pulse amplitude is 0.1 mT and its temporal dependence is of type 
$\exp(- a(t-t_0))$ for t > t$_0$ and 0 elsewhere. The parameter $a$ was set to 0.5 ns$^{-1}$. The damping parameter is reduced from $\alpha$ = 1 used for the static simulations to a smaller value of 0.08 for NdCo$_5$ and 0.01 for Py \cite{Zhao2016}. After the pulse was applied, the total magnetization has been evolved for 20 ns and the average out-of-plane magnetization of Py was recorded. Its value was fast Fourier transformed and the frequency of the resulting peak is determined as $H$ is varied.
\begin{figure}[h]
 \centering
  \includegraphics[width=0.49\textwidth]{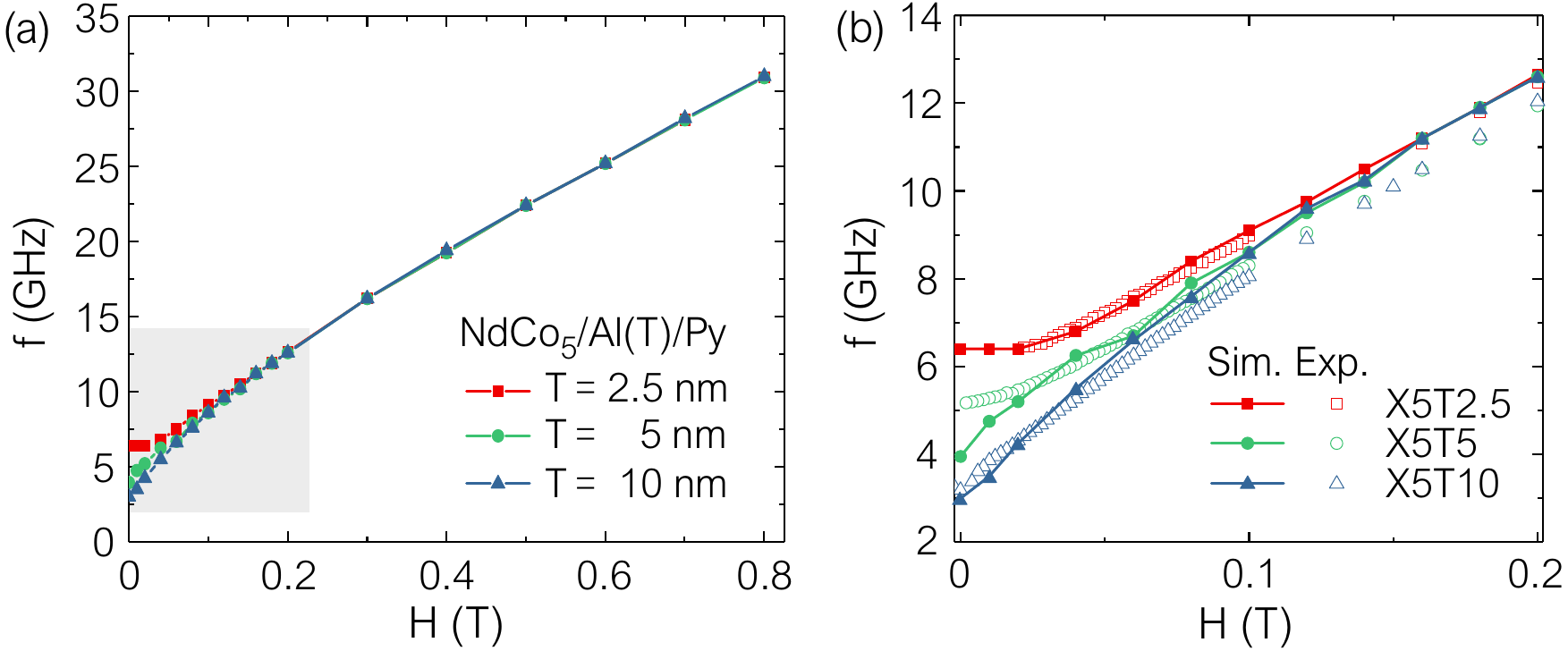}
	 \caption{(Color online) (a) Simulated FMR frequencies of the Py film in NdCo$_5$-based trilayers with varying Al spacer thickness $T$ for a positive, decreasing in-plane magnetic field $H$ applied along the 
						HA.	(b) Comparison between simulated (filled symbols) and measured (open symbols) resonance frequencies within the gray-shaded area of (a).}
	\label{fig:Fig3}
\end{figure}

In Fig.~\ref{fig:Fig3}(a), the $f$ vs.~$H$ dependence of the Py film in a NdCo$_5$-based trilayer for a positive, decreasing in-plane magnetic field applied along the HA is shown. In excellent agreement with 
experimental results, the increased zero-field resonance frequencies as well as the overall applied field and Al spacer thickness dependency are reproduced. From the overlay of experimental and simulated data in the 
frequency and field range marked by the gray-shaded area in Fig.~\ref{fig:Fig3}(a), depicted in Fig.~\ref{fig:Fig3}(b), it can be seen that there is generally a very good quantitative agreement, except that the 
simulations yield a 1 GHz higher zero-field frequency for the X5T5 sample as well as slightly increased values of $f$ for the X5T5 and X5T10 trilayers at higher fields.

\section*{References}
\bibliography{OviedoAPLwithoutTitles}